\documentclass[sigconf]{acmart}
\AtBeginDocument{%
  }

\copyrightyear{2026}
\acmYear{2026}
\setcopyright{cc}
\setcctype{by}
\acmConference[ICSE-FoSE]{2026 IEEE/ACM 48th International Conference on Software Engineering: Future of Software Engineering }{April 12--18, 2026}{Rio de Janeiro, Brazil}
\acmBooktitle{2026 IEEE/ACM 48th International Conference on Software Engineering: Future of Software Engineering (ICSE-FoSE), April 12--18, 2026, Rio de Janeiro, Brazil}
\acmPrice{}
\acmDOI{10.1145/3793657.3793901}
\acmISBN{979-8-4007-2479-4/2026/04}

\newcommand{\topic}[1]{\textbf{#1.}}

\begin{document}

\title[Awakening]{Awakening: Modern Challenges and Opportunities\\
of Software Engineering Research}

\author{Diomidis Spinellis}
\orcid{0000-0003-4231-1897}
\affiliation{%
 \institution{Athens University of Economics and Business}
  \city{Athens}
  \country{Greece}
}
\email{dds@aueb.gr}
\affiliation{%
  \institution{Delft University of Technology}
  \city{Delft}
  \country{Netherlands}
}

\author{Zoe Kotti}
\orcid{0000-0003-3816-9162}
\affiliation{%
 \institution{Athens University of Economics and Business \&}
  \city{}
  \country{}
}
\affiliation{%
 \institution{DeepSea Technologies}
  \city{Athens}
  \country{Greece}
}
\email{zoekotti@aueb.gr}

\renewcommand{\shortauthors}{Spinellis \& Kotti}

\begin{abstract}
Software engineering research benefited for decades from openly available tools,
accessible systems, and problems that could be studied at modest scale.
Today, many of the most relevant software systems are large, proprietary,
and embedded in industrial contexts that are difficult to access or
replicate in academia.
We review how the field reached this point,
identify structural challenges facing contemporary research,
and argue that incremental methodological refinement is insufficient.
We discuss practical directions forward, including industrial PhDs,
long-term industry-academia collaborations,
larger research teams, moonshot projects, and changes to funding and
evaluation practices.
\end{abstract}



\maketitle

\section{Introduction}
In \emph{Awakenings}~\cite{Sac73}, neurologist Oliver Sacks reports on his work
at a New York hospital treating patients with post-encephalitic
Parkinsonism who had spent decades in extreme rigidity, mutism,
and institutional stasis.
The book presents detailed case histories of patients given L-DOPA
in the late 1960s, a treatment that at first restored movement, speech,
and initiative with striking clarity.
These sudden awakenings were short-lived: dosage escalation
led to severe side effects, emotional instability,
and the gradual return of rigidity or new pathological states,
forcing patients and clinicians
to confront the limits of pharmacological recovery.
\emph{Awakenings}
thus records not a cure but a transient reopening of consciousness.

Here we argue that software engineering research has passed though a similar
temporary happy age of innocence where development processes, projects,
tools, and methods allowed the advancement of knowledge in a simple,
effective, open, and reproducible manner.
A victim of its own success,
software engineering research is now conducted in an environment
where the complexity of processes and artifacts is beyond what can be
easily developed or studied in an academic laboratory.
Researchers often have to choose between
relatively trivial studies of unrealistic toy problems and
multidimensionally challenging studies in proprietary contexts.

\section{A Brief History of Software Engineering Research} 
The following paragraphs trace software engineering research
from its inception to the present,
illustrating how early productivity driven by
openly available tools, low-hanging fruit, and open-source software
gave way to Big Tech's research advantages.

\topic{Birth in a Crisis} 
Software engineering emerged in response to a crisis of
chronic overruns, unreliability, and escalating maintenance costs.
The 1968 NATO Software Engineering Conference~\cite{NR68}
crystallized this diagnosis and established an engineering framing
focused on disciplines, methods, and measurement.
The community institutionalized through dedicated conferences,
starting with the 1975 meeting on reliable software~\cite{ICRS75}
(often called ``ICSE-0''~\cite{BMR18}),
where 22 of 53 papers came from industry and four from government.
At the time, the raw material needed for rigorous research---large
codebases, defect reports, effort data, tools, and operational experience---lived
inside companies and was largely proprietary.
This made early software engineering research unusually dependent
on industrial access and anecdote-driven generalization.

\topic{Tool Building} 
A defining force in early software engineering research
was building tools for software development tasks.
AT\&T Bell Laboratories developed and published
foundational technologies including
filters~\cite{Meu95},
file differencing and archives,
pattern matching,
macro processing,
language translation~\cite{KP76},
programmer's workbenches~\cite{DHM78},
language development tools~\cite{JL78},
build automation~\cite{MAKE},
configuration management~\cite{SCCS},
static analysis~\cite{LINT}, and
domain-specific languages~\cite{Ben86b}.
Importantly, a 1956 consent decree~\cite{Lew56} prohibited AT\&T from
commercializing non-telecommunications products,
so Unix and its tools were liberally licensed royalty-free~\cite[p. 60]{Sal94}
with widely available source code.
This allowed universities worldwide to apply, study,
and enhance these tools.

\topic{Low-Hanging Fruit} 
Pinning down known-unknowns allowed low-hanging fruit to be collected.
One strand concerned
the tighter management of the software process.
For example,
\citet{Bro75} performed early empirical action research on large-scale
software development,
while \citet{Hum89} introduced
software process maturity models and levels,
process assessment and change,
and several software process elements we now take for granted:
the project plan,
configuration management
quality assurance,
software inspections and testing,
data gathering and analysis,
automation and defect prevention.
Other influential researchers contributed to the economics of software
development and its maintenance~\cite{Boe81,Boe83}
and the understanding of software evolution~\cite{Leh96}.
On the product side,
major advancements concerned
structured programming~\cite{Dij68b},
axiomatic program correctness~\cite{Hoa69},
information hiding~\cite{Par72},
program families~\cite{Par76},
and model checking~\cite{CES09}.

\topic{Open Source Software} 
The emergence of open-source software~\cite{ASKG10,Fit06}
was a boon for software engineering research.
First, it provided an avenue for disseminating and evaluating
new software tools, ranging
from configuration management systems~\cite{RCS,CS14},
to unit testing frameworks~\cite{BG98},
to static analysis tools~\cite{AHMP08}.
Second,
it provided product and process data of key software,
such as the Linux kernel~\cite{SJWH02},
Eclipse~\cite{YWA05}, the Apache web server~\cite{WH05},
and MySQL~\cite{HCY19},
with repositories that could be easily and profitably analyzed
at scale~\cite{KH06,BRBH09,KGBS16}.
Third, it advanced software reuse,
an early key software engineering goal~\cite{MBNR68,CB91},
through tens of thousands freely available (easily and a zero or little cost)
open source components~\cite{Spi07a,HKS08}
efficiently distributed and deployed
through package management systems~\cite{Spi12b}.

\topic{Big Tech} 
Over recent decades, network effects~\cite{KS94}
enabled by internet technologies~\cite{GW02}
provided virtual exclusivity to Big Tech
in impactful software engineering research
through control of the modern experimental substrate:
hyperscale systems, telemetry, and real-world interventions at scale.
Research within platforms
(cloud, app stores, browsers, IDEs, CI, repo hosts)
can be evaluated and shipped faster and more effectively
than university prototypes.
Frontier areas (ML systems, LLM tooling, security, reliability)
are hard to study without industrial workloads and infrastructure,
while tools are often provided opaquely through SaaS offerings,
hindering open study.
Recent research shows sustained Big Tech research growth
and deep university co-authorship ties
that concentrate agenda-setting power~\cite{MSD25}.

Network effects amplify this trend in two ways.
First, collaboration networks:
once a few firms become hubs for high-impact work,
top researchers and students preferentially connect to them
(co-authorship, internships, datasets),
creating cumulative advantage~\cite{LZZC22}.
Second, platform network effects~\cite{BP18}:
when software is built on dominant platforms
(GitHub, major cloud providers),
the platform owner controls instrumentation, access, and integration.
This makes impactful research disproportionately
the kind that rides those platforms' distribution and data loops,
while alternatives struggle to reach the same scale~\cite{CLC16}.
Consequently, key software artifacts and processes
are again becoming off-limits to researchers,
returning the field to its starting point.

\section{Modern Challenges} 
Contemporary software engineering research faces
qualitatively different challenges.
While foundational questions
have largely been articulated,
today's most consequential problems are embedded in sociotechnical systems
that are larger, more opaque, and more tightly coupled
to industrial practice.
These shifts push impactful research toward problems
that resist traditional academic methods,
strain PhD-scale projects,
and raise hard questions about access, scale, and legitimacy.

\topic{Action Away from the Streetlight} 
As low-hanging fruit have been picked by previous generations of
software engineering researchers, the interesting and consequential software
research is no longer near the proverbial streetlight
where the drunk searches for his keys because ``that's where the light is''.
Yet, this principle of the drunkard's search~\cite{Kap17} is often
seen in published software engineering research:
polished and methodologically sound papers exploring phenomena that are
relatively easy to research but of little potential practical or
theoretical impact.
We fear that such research will become even more common with the aid of
generative AI systems, which excel at suggesting appropriate methods
and churning out word-perfect prose, but often lack
in tackling really hard and consequential problems.

\topic{Scale and Complexity} 
\label{sec:scale-complexity}
Modern software development, from public-sector contracts to Big Tech efforts,
characteristically resists study by a lone PhD
student --- a typical structure of software engineering research.
Specifically,
both types of software development often occur over long timescales
that do not fit the typical 3--4 year PhD program.
Furthermore, they occur at a scale of organizational and technical complexity
that is difficult for an inexperienced PhD student to grasp and tackle
in a profitable manner.
Focusing the research on a particular element of software development
can be problematic, due to the risk of ignoring interdependencies and
their effects.
As an example regarding the scale of modern commercial software development
Google reports two billion lines of code in nine million unique source files,
one billion files, 35 million commits, more than 25 thousand developers and about 15 million lines of code changed weekly~\cite{PL16}.
Although this data set would be a goldmine for software engineering researchers,
its proprietary value and size place it beyond the reach of most PhD students.

\topic{Cloud Computing and Foundational Models} 
Modern software development increasingly relies on opaque cloud offerings~\cite{SOKO24}
and fickle foundational models~\cite{SDP24},
creating significant barriers to rigorous research.
Cloud platforms operate as black boxes with proprietary implementations~\cite{FMTI25},
undocumented behavior~\cite{Kha25}, and ever-changing APIs~\cite{CJER22}.
Researchers face opaque infrastructures~\cite{KAN24},
difficulties in controlling experimental conditions~\cite{CJER22},
challenges in replicating results across providers~\cite{KAN24} or time~\cite{CJER22},
prohibitive costs~\cite{Bor24}, and
vendor lock-in that constrains generalizability~\cite{SOKO24, KAN24}.

Foundational models present similar challenges amplified by their scale.
These models are frequently updated, deprecated, or discontinued without notice,
making longitudinal studies nearly impossible~\cite{CZZ24}.
Their outputs are non-deterministic and version-dependent~\cite{CZZ24, PELH23},
undermining reproducibility~\cite{SDP24},
which is a cornerstone of modern science~\cite{Rep19}.
Access through rate-limited APIs rather than downloadable artifacts
prevents independent verification~\cite{PELH23}.
Training data, architectures, and fine-tuning remain largely proprietary,
reducing researchers to studying systems they cannot fully understand~\cite{SDP24, FMTI25}.
Together,
cloud computing and foundational models introduce opacity
that contrasts starkly with open-source ideals~\cite{Ray99},
forcing research to grapple with increasingly opaque and ephemeral topics.

\topic{Publication Culture} 
The publication culture of software engineering research increasingly
shapes how research is conducted,
often in counterproductive ways.
Strong pressure on PhD students and early-career researchers to produce
multiple publications within a short time frame incentivizes shallow,
incremental work over sustained investigation.
Techniques such as salami slicing, excessive parameter variation,
and the study of marginal or streetlight research
become rational responses to evaluation criteria that reward paper counts,
venue prestige,
and methodological polish rather than depth, relevance,
or long-term impact~\cite{LNZ15}.
As a result, research questions are often chosen for publishability
rather than importance, and negative results, replication,
and long-horizon studies are systematically disfavored.

These dynamics are reinforced by the structure of the software engineering
research community.
For a field of a few thousand active researchers,
the existence of multiple flagship and top conferences,
alongside a growing ecosystem of adjacent venues,
fragments contributions and amplifies
incentives to split work across outlets~\cite{VSMv14}.
This venue proliferation,
apart from its impact on CO$_2$ emissions~\cite{SL13},
promotes triviality,
inflates review workloads~\cite{SKP20}
to levels uncommon in neighboring disciplines,
and reduces impact evaluation to venue prestige and publication volume.

\section{The Way Ahead} 
The challenges outlined in the previous section are structural
rather than accidental, and cannot be addressed by methodological
refinement alone.
Responding requires rethinking how software engineering research
is organized, funded, and evaluated, and how it interfaces
with industrial practice.
We outline concrete directions for increasing relevance,
feasibility, and impact,
discussing industrial research models,
coordinated efforts,
moonshot projects, and
funding structures required to sustain progress
in the face of growing scale and complexity.

\topic{Industrial Research} 
Doctoral education is moving beyond academic careers~\cite{OB23,Cab01},
with industrial PhD collaborations emerging
where students work at firms while pursuing degrees.
These dual-affiliated students bridge academia and industry~\cite{BMS25},
bringing practical relevance and academic rigor~\cite{Bos24}.
As LLMs enable drafting scientific publications in hours~\cite{LWMN23},
industrial PhDs emphasizing real-world problems~\cite{MTB11}
may offer a more resilient model.
Software engineering research must embrace these partnerships
for greater impact.

Publication venues can bridge research and practice.
Practitioner-oriented venues such as
\emph{Software: Practice and Experience} and \emph{IEEE Software}
demonstrate higher industrial impact than researcher-oriented venues~\cite{KGS23}.
Conferences with practitioner tracks---ICSE's \emph{Software Engineering in Practice}
and \emph{Tool Demonstrations} and \emph{Industry Showcase} at FSE and ICSME---rank
considerably higher in patent citations.
Venues should cultivate industrial engagement
through dedicated calls and tracks addressing real-world problems.

\topic{Industrial Collaborations} 
Setting up a single industrial PhD is cumbersome and inefficient for
all parties involved.
For the academic faculty it often requires protracted negotiations
with the hosting company and the university to find a mutually
agreeable collaboration framework.
The hosting company also needs to secure the required management
commitment and set up an effective collaboration framework.
On top of that comes the effort required to find and hire a PhD student
that fits both organizations.

Large-scale collaboration frameworks where industrial partners
fund specialized laboratories hosting multiple PhD students
amortize these overheads across students.
For example,
TU Delft has established such collaborations with
ING (on FinTech~\cite{KCSv24}),
JetBrains (on AI4SE~\cite{Pan25}), and
Meta Platforms (on the future of software engineering).
Following established best practices~\cite{GPFF19,MG21}
helps overcome well-known challenges~\cite{Woh13}.

\topic{Large-Scale Collaborations} 
The research community should actively encourage
and reward larger-scale collaborations
that bring together complementary expertise, resources, and
perspectives~\cite{SMSW17}.
Such collaborations can tackle problems of greater scope and complexity,
produce more robust and generalizable solutions, and
accelerate the path from research to practice~\cite{WAAP12}.
However,
larger collaborations introduce legitimate concerns about attribution and
fairness in assigning credit~\cite{AABB25}.
These are particularly critical
for early-career researchers whose academic progression depends
on demonstrating individual contributions~\cite{Lee23}.

Structured contribution declaration systems,
such as the ICMJE authorship criteria~\cite{ICMJE}
and the CRediT (Contributor Roles Taxonomy)~\cite{CRediT} framework,
offer a solution to this challenge.
CRediT, which was used in over 20\% of scholarly publications in 2024~\cite{AKPW25},
provides a standardized vocabulary of 14 contributor roles---including
conceptualization, methodology, software development, validation, and writing---that
enables fine-grained, transparent attribution of individual contributions
within collaborative work.
By making explicit who contributed what to a research project,
such systems can support larger collaborations
while maintaining accountability and
enabling fair evaluation of individual researchers~\cite{AKPW25}.

\topic{Moonshot Projects} 
Another way to increase the possibility of impact is
moonshot projects~\cite{CF16,MPT19}.
These are projects comparable to the scope and ambition of the
1960 Apollo spaceflight program, which landed the first humans on the Moon.
In the case of software engineering research one can mention
\citeauthor{Hoa03}'s \citeyear{Hoa03}
proposal  for a formally validated compiler~\cite{Hoa03}
and the more recent US DARPA project aiming to automatically
convert all C code into safe Rust~\cite{TRACTOR}.
Projects of similar ambition could be for technologies that
can close correctly 90\% GitHub issues through generative AI
or to increase the throughput and efficiency of test runs
by orders of magnitude while maintaining their effectiveness.

Moonshot projects force step-function advances by relaxing
incremental constraints, making it rational to pursue high-risk
ideas that would be rejected under normal cost–benefit analysis.
Even when they fail, they generate durable spillovers:
new tools, methods, and trained people.
These often reshape adjacent fields and reset what is considered feasible.

\topic{Funding} 
Funding agencies must come to terms with the new landscape and
realize that toy funding for solving toy problems is a waste of
human and financial resources.
Recognizing that software engineering research results can
raise their country's economic performance,
they should increase funding to the levels needed for advancing
modern software engineering and the industry's competitiveness~\cite{Spi17d},
while also encouraging the required industrial collaborations.

Companies can leverage their resources (data, problems, developer access,
infrastructure) through collaborations with academic researchers.
While companies lose some control compared to in-house research,
collaborations bring new ideas, allow focus on strategic goals,
and are more cost-effective.
Professional societies, such as the ACM and IEEE, can recognize
collaborative industrial research
through awards,
while governments can incentivize it through dedicated funding and tax benefits.

\topic{Publications and Incentives} 
On the publication side,
a smaller number of conferences and journals,
combined with higher selectivity and a greater tolerance for non-publication,
would shift incentives toward fewer, deeper, and more consequential
contributions.
This would align publication practices more closely
with the realities and needs of modern software engineering.

\section{Conclusions} 
The sweet age of innocence is over.
Software engineering research can no longer rely on the
fortuitous alignment
of open artifacts, tractable systems, and observable processes
that characterized its early productive and impactful decades.
The field now operates amid scale, opacity,
proprietary control, and sociotechnical entanglement,
where consequential systems are inaccessible
or irreproducible academically.
This is structural change, not methodological failure:
as software became the substrate of economic activity,
resources to study it concentrated in industry,
while academic incentives remained unchanged.
The mismatch risks pushing research
toward polished but marginal problems.
Addressing this requires
rethinking how research is conducted,
not merely evaluated.
Closer industry integration,
larger coordinated efforts,
recognition of collaborative contributions,
and moonshot projects
are necessary adaptations.
Funding agencies, professional societies,
and publication venues must evolve
to reward relevance and impact
alongside rigor.
To remain a driver of progress
rather than a retrospective observer,
the field must accept the loss of its former simplicity
and rebuild around today's realities.
Like the awakenings described by Sacks,
early clarity was real but transient.
A next awakening depends on the choices our community now makes
about scale, access, collaboration, and ambition.

\bibliographystyle{ACM-Reference-Format}
\IfFileExists{used.bib}
  {\bibliography{used}}
  {\bibliography{macro,bib,ddspubs,myart,classics,coderead,unix,various,mybooks,bigdata,ieeestd,isostd}}

\end{document}